\documentclass[journal]{IEEEtran}

\ifCLASSINFOpdf

\else

\fi
\usepackage{amssymb}
\usepackage{xcolor}
\usepackage{cite}
\usepackage{multirow}
\usepackage{booktabs}
\usepackage{makecell}
\usepackage{amsmath}
\usepackage{textcomp}
\usepackage{bm}
\usepackage{color}
\usepackage{setspace}
\usepackage{amsmath}
\usepackage{amsthm,amsmath,amssymb}
\usepackage{mathrsfs}
\usepackage{graphics}
\usepackage{array,diagbox,siunitx}
\usepackage{graphicx}
\usepackage{subfigure}
\usepackage{url}
\usepackage{hyperref}
\usepackage[hyphenbreaks]{breakurl}

\usepackage{epsfig}
\usepackage{amssymb}

\usepackage{lineno,hyperref}
\modulolinenumbers[5]

\begin{document}
\title{{Semantic Importance-Aware Communications  Using   Pre-trained Language Models}}

\author{Shuaishuai Guo, \emph{Senior Member}, \emph{IEEE}, Yanhu Wang,  Shujing Li, and Nasir Saeed, \emph{Senior Member}, \emph{IEEE}
\thanks{The work is supported in part by the National Natural Science Foundation of China under Grant 62171262; in part by Shandong Provincial Natural Science Foundation under Grant ZR2021YQ47; in part by the Taishan Young Scholar under Grant tsqn201909043; in part by Major Scientific and Technological Innovation Project of Shandong Province under Grant 2020CXGC010109. (\emph{*Corresponding author: Shuaishuai Guo}).}
\thanks{S. Guo, Y. Wang, and S. Li are  with the School of Control Science and Engineering, Shandong University, China, and also with Shandong Key Laboratory of Wireless Communication Technologies, Shandong University, China (E-mail:shuaishuai$\_$guo@sdu.edu.cn; yh-wang@mail.sdu.edu.cn; lishujing777@mail.sdu.edu.cn ). 
N. Saeed is with the Department of Electrical and Communication Engineering, United Arab Emirates University (UAEU), Al Ain, United Arab Emirates. E-mail: mr.nasir.saeed@ieee.org.}
}

\maketitle

\begin{abstract}
This letter proposes a semantic importance-aware communication (SIAC) scheme using pre-trained language models (e.g., ChatGPT, BERT, etc.). Specifically, we propose a cross-layer design with a pre-trained language model embedded in/connected by the cross-layer manager. The pre-trained language model is utilized to quantify the semantic importance of data  frames. Based on the quantified semantic importance,  we investigate semantic importance-aware power allocation. Unlike existing deep joint source-channel coding (Deep-JSCC)-based semantic communication schemes, SIAC can be directly embedded into current communication systems by only introducing a cross-layer manager. Our experimental results show that the proposed SIAC scheme can achieve lower semantic loss than existing equal-priority communications.
\end{abstract}


\IEEEpeerreviewmaketitle

\section{Introduction}
Semantic communications aim to convey semantic information efficiently and minimize semantic-level loss rather than bit-level errors during transmission \cite{Qin2021Survey}. There exist three different kinds of applications that urgently need semantic communications. The first type requires large amounts of data to be transmitted in a short period, such as augmented reality/virtual reality (AR/VR), intelligent transportation/factory, and healthcare \cite{9955525}. The second type consists of human-machine communications, as human beings communicate efficiently at a semantic level and wish to communicate with machines as well at the semantic level. The third type is communication in harsh environments where the transmission suffers from severe channel attenuation, such as deep space communications \cite{9952925} and underwater communications \cite{6605598}. 

Semantic communications are currently implemented using two different ways. The first way is the traditional source-channel separation method, where the transmitters deeply compress the data and then transmit the semantic information across the channel. In this way, semantic communications only involve source compression and decompression to minimize semantic loss. Such an approach is compatible with existing communication systems with block designs by only introducing semantic encoding, and decoding \cite{9791398, niu2022towards}. However, in such designs, the source and channel parts are totally isolated, which may be far from the joint optimal design.
The second way of implementing semantic communications is joint source-channel coding (JSCC) designs by exploiting the potentials of deep learning (DL) \cite{farsad2018deep, Xie2020,10012981,bourtsoulatze2019deep, Wang2023T}. 
For instance, \cite{farsad2018deep} utilized the bidirectional long and short term memory network (BLSTM) as JSCC for text transmission to minimize the word error rate. Then, \cite{Xie2020} proposed a DL method for JSCC to minimize semantic level loss, referred as Deep-JSCC.
\cite{10012981} recently defined a semantic loss measured by the pre-trained bidirectional encoder representation from transformers (BERT) model and proposed a signal shaping method to minimize the semantic loss. The interested readers are referred to \cite{Lan2021what} and \cite{luo2022semantic} for recent progress in semantic communications.

Despite the impressive performance of semantic communication schemes based on the JSCC architecture, especially at low signal-to-noise ratio (SNR) regimes, some issues are still worth further discussion.
Firstly, due to the limitations of DL technology, universal semantic-level JSCC techniques have yet to be available \cite{Qin2021Survey}, i.e., the transceiver needs to be re-designed and re-trained if the task at the receiver or the experienced channels changes. Also, when the network model of the transceiver is large, the cost of training becomes unaffordable.
Secondly, guided by Shannon’s source-channel separation theorem \cite{Shannon1948},
existing communication systems mainly adopt separate source and channel coding designs and layered structures. These semantic communication schemes based on the JSCC architecture are incompatible. Therefore, if existing communication systems want to adopt these semantic communication technologies, it would require updating all devices and systems, resulting in high costs.

To avoid the above issues, this letter investigates semantic importance-aware communication schemes (SIAC) {by only introducing a cross-layer manager to existing communication systems}. Inspired by human-to-human semantic communications where one may speak the \emph{important} words louder (i.e., using more power) or repeat \emph{important} words by many times (i.e., repetition coding), we are interested in {quantifying the semantic importance and developing SIAC} systems. A key question is how to recognize the \emph{important} words as that in human-to-human communications.
It is challenging to develop a mathematical model to quantify the semantic importance, as it is related to the context, background knowledge, and many other factors. 

\begin{figure*}
\centering 
    \includegraphics[width=0.73\linewidth]{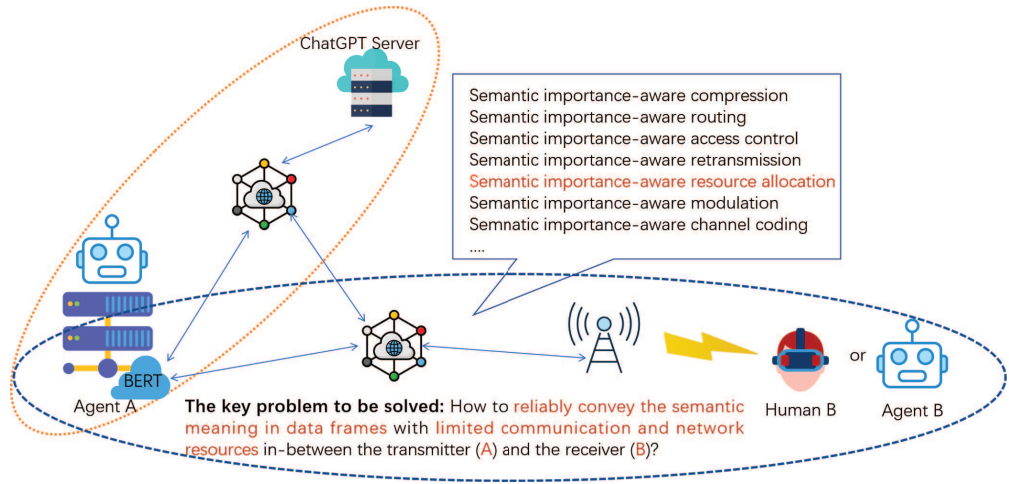}
    \caption{Demonstration of SIAC systems, which target to reliably transmit the semantic meaning in data frames.}
\label{system_model_LLM}
\end{figure*}

In this letter, we utilize pre-trained language models  to quantify the semantic importance of frames. Specifically, two methods are proposed.  The first method adopts pre-trained generative language models. We take the representative chat generative pre-trained transformer (ChatGPT) as an example, and the corresponding communication scheme is referred as ChatGPT-SIAC. It is developed for supporting reliable cloud chat robot (CCR)-to-human semantic communications, where the messages to be transmitted are generated by ChatGPT. In such applications, ChatGPT is only required to emphasize the important words while generating the communication content. 
However, it is noteworthy that  ChatGPT-SIAC needs access to the closed-source ChatGPT\footnote{As of March 1st, 2023, the license for using ChatGPT is subject to the OpenAI Platform License v1.0. Under the license, one is granted access to the ChatGPT and is permitted to use it to interact with the model.} and suffers from the randomness of the generative language model. To handle these challenges and broaden the application scenarios of SIAC, we further propose another semantic importance analysis method using pre-trained discriminative language models. We take the representative open-source BERT\footnote{BERT is released under the Apache License 2.0, which is a permissive open-source license. The Apache License 2.0 allows one to use, modify, and distribute BERT for both commercial and non-commercial purposes. It grants you the freedom to use the software without restrictions, subject to certain conditions outlined in the license.} model as an example and refer to the transmission scheme as BERT-SIAC.
Based on the semantic importance of frames output by ChatGPT or BERT, the transmitter can then implement SIAC strategies, making the transmission of important frames more reliable. 
Our experimental results show that  the proposed ChatGPT-SIAC and BERT-SIAC can considerably outperform the existing equal-priority communication strategy in terms of semantic loss.

\section{System Model}

\begin{figure}
       \centering  
       \includegraphics[width=0.72\linewidth]{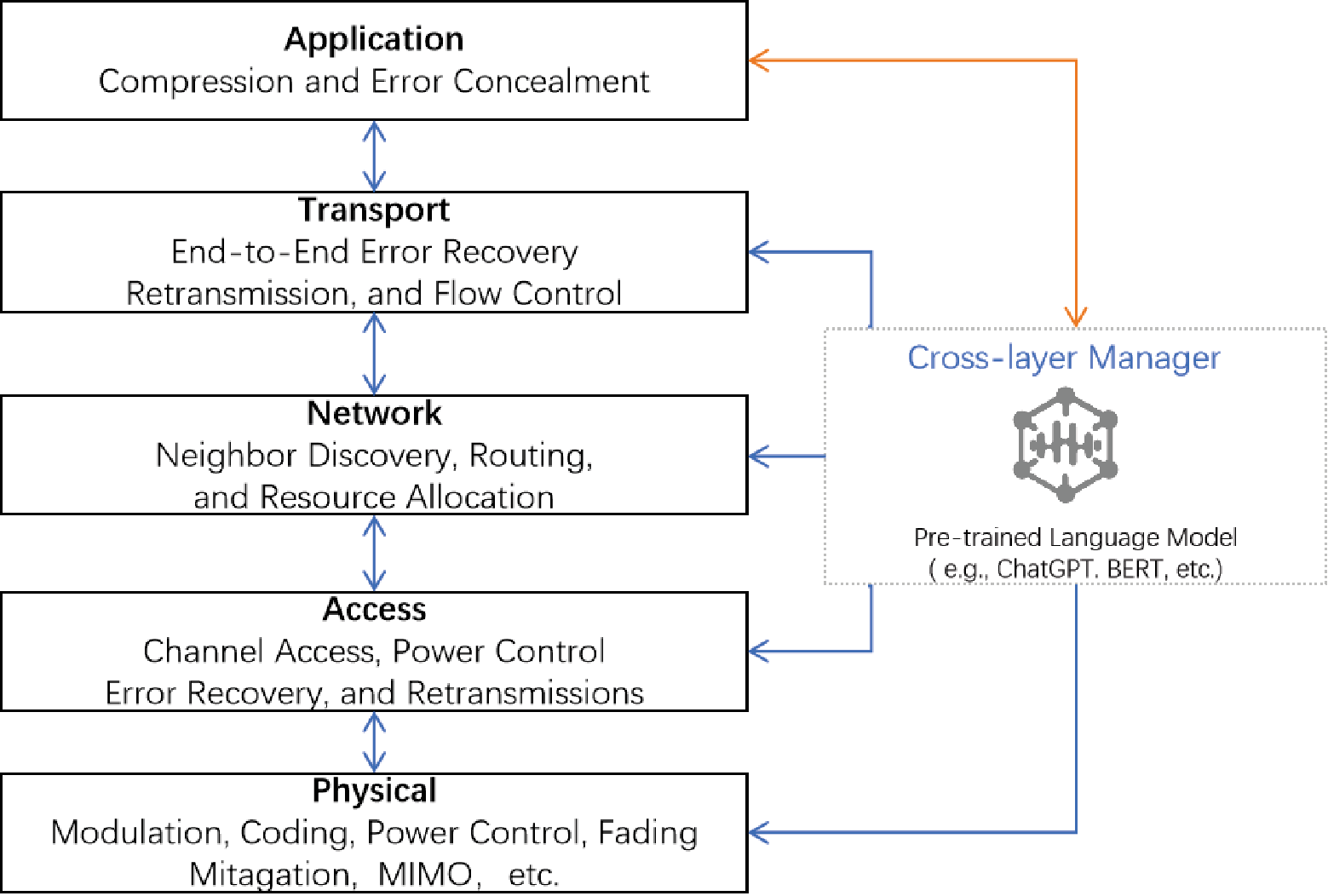}
       \caption{SIAC cross-layer structure with pre-trained language model (e.g., ChatGPT, BERT, etc.) embedded in/connected by the cross-layer manager.}
       \label{Transmitter}
\end{figure}

In this letter, we investigate a communication system that  targets to reliably transmit semantic meaning in data frames with limited communication and network resources in-between the transmitter and the receiver, as illustrated in Fig. \ref{system_model_LLM}.
The potential applications are CCR-to-human communications, or agent-to-agent semantic communications. At the transmitter, we assume there is a batch of messages to be transmitted. In the case of CCR-to-human communications, the messages may be generated by generative language models, e.g., ChatGPT, or selected from a knowledge base system. Before transmission, these messages are split into $N$  number of total frames. Conventional  data-oriented communication systems usually overlook the semantic meaning of the data to be transmitted. 
Therefore, in this letter, we aim to minimize the semantic loss during the data transmission.
To do so, we quantify the importance of each frame and perform semantic importance-aware priority-based communications for each frame. Specifically, we propose a cross-layer design with a pre-trained language model embedded in/connected by the cross-layer manager as illustrated in Fig. \ref{Transmitter}. The manager is connected to all layers and by interacting with the application layer, the pre-trained language model can output  the semantic importance of frames. By passing the frame importance to the physical, access, network, transport, and application layers, priority-based parameter configurations can be conducted in one or multiple layers. Here, we list some of the priority-based communication technologies that can be adopted:
\begin{table*}[t]
\footnotesize
\caption{Important ('1') and Non-Important ('0') Words in   ``It is an important step towards equal rights for all passengers''.}
\label{gpt}
\centering
\renewcommand{\arraystretch}{1.5}
\begin{tabular}{|c|c|c|c|c|c|c|c|c|c|c|c|c|}
\hline
\textbf{Frame Data} & It &is &an &important &step &towards &equal &rights &for &all &passengers&\textbf{Frame Importance}\\
\hline
\textbf{ChatGPT Output 1} & 0 & 0 & 0 & 1& 1 & 0 & 1 & 1 & 0 & 1 & 1&6\\
\hline
\textbf{ChatGPT Output 2} & 0 & 0 & 0 & 1& 1 & 0 & 1 & 1 & 0 & 1 & 1 &6\\
\hline
\textbf{ChatGPT Output 3} & 0 & 0 & 0 & 1& 1 & 0 & 1 & 1 & 0 & 1 & 1&6\\
\hline
\textbf{ChatGPT Output 4} & 0 & 0 & 0 & 1& 1 & 0 & 1 & 1 & 0 & {0} & 1&{5}\\
\hline
\textbf{ChatGPT Output 5} & 0 & 0 & 0 & 1& 1 & 0 & 1 & 1 & 0 & 1 & 1&6\\
\hline
\textbf{Majority Vote} & 0 & 0 & 0 & 1& 1 & 0 & 1 & 1 & 0 & 1 & 1&6\\
\hline
\textbf{Arithmetic Average} & 0 & 0 & 0 & 1& 1 & 0 & 1 & 1 & 0 & 0.8 & 1&5.8\\
\hline
\end{tabular}
\end{table*}
\begin{itemize}
    \item
    \textbf{Physical layer:} Importance-aware modulation, coding, power control, fading mitigation, and multiple-input multiple-output (MIMO) setups;
     \item 
    \textbf{Access layer:} Importance-aware channel access, power control, error recovery, and retransmissions;
    \item
    \textbf{Network layer:} Importance-aware routing and resource allocation;
     \item
    \textbf{Transport layer:} Importance-aware error recovery, retransmissions, and flow control;
    \item 
    \textbf{Applications layer:} Importance-aware compression and error concealment.   
\end{itemize}
The key to realizing SIAC is recognizing the semantic importance of a frame.  However, it is challenging to develop a mathematical model to quantify the semantic importance, as it is related to the context, background knowledge, and many other factors. In the following section, we will present our methods to quantify the semantic importance using pre-trained  language models.

\section{Semantic Importance Quantification}
In this work, we propose two methods to quantify the semantic importance of frames using pre-trained language models. The first one uses pre-trained generative language models. For demonstration, we take the representative ChatGPT as an example.  The second one relies on pre-trained discriminative language models 
and we take the representative BERT model as an example.
\subsection{Using Generative Language Models Like ChatGPT}
In CCR-to-human communications, the current most advanced and widely used chat robots are relying on the access of ChatGPT.
As the communication content may  be generated by ChatGPT, it is straightforward to directly use ChatGPT for recognizing the semantic importance of its generated content. To quantify the semantic importance of frames, we define the frame importance to be the number of words it contains as illustrated in Table \ref{gpt}.
The more important words a frame contains, the more important the frame is. Based on this definition, ChatGPT only needs to emphasize the important words when generating the communication content. By counting the important words in a frame, the semantic importance of the frame can be quantified. 

Nevertheless, this method faces two major challenges. The first challenge is the requirement of accessing the closed-source ChatGPT, narrowing the scope of its usage. The second challenge is that ChatGPT as a generative model has inherent randomness, which can be observed by the fourth output 
of ChatGPT in the example shown in Table \ref{gpt}. This implies that the critical words ChatGPT recognizes for the same content may differ. As the word importance recognition is done before transmission, it can be conducted often. The randomness can be minimized by fusing multiple results via majority vote or arithmetic average, as shown in Table \ref{gpt}.

\subsection{Using Discriminative Language Models Like BERT}
Motivated  by broadening the application scenarios of SIAC and avoiding the randomness of a generative language model, we further propose an importance indicator of a word/frame using the pre-trained discriminative language models like BERT.
Specifically, we define the word/frame importance using the semantic loss caused by the missing word/frame.  More loss indicates that the word/frame is more important to the semantic meaning. In this paper, we adopt the semantic loss defined in our previous work \cite{10012981} using the pre-trained BERT model, which is expressed as
\begin{equation}
\label{Loss}
L(\mathbf{m}_i,\mathbf{m}_j)=1-\phi(\mathbf{m}_i,\mathbf{m}_j),
\end{equation}
where $\phi(\textbf{m}_i,\textbf{m}_j)=\frac{B_{\pmb{\psi}}(\mathbf{m}_i)^TB_{\pmb{\psi}}(\mathbf{m}_j)}{||B_{\pmb{\psi}}(\mathbf{m}_i)||\cdot|| B_{\pmb{\psi}}(\mathbf{m}_j)||} $ represents the semantic similarity of two messages $\mathbf{m}_i$ and $\mathbf{m}_j$.
$B_{\pmb{\psi}}(\cdot)$ stands for the pre-trained BERT model.

To demonstrate the idea, we list an example in Table \ref{semloss}.
Intuitively, for the sentence ``It is an important step towards equal rights for all passengers'', words like ``It'', ``is'' and ``for'' are not important. When ``step'', ``rights'', and `` passengers'' are missing, it may change the semantic meaning of the sentence.
Table \ref{semloss} shows that ``step'', ``rights'', and ``passengers'' have a more semantic loss, indicating their higher importance.

\begin{table}[t]\footnotesize
\caption{Semantic Loss of Each Word in the Sentence  ``It is an important step towards equal rights for all passengers''}
\label{semloss}
\centering
\renewcommand{\arraystretch}{1.5}
\begin{tabular}{|c|c|c|c|}
\hline
\textbf{Word} & \textbf{Semantic loss}   & \textbf{Word}          & \textbf{Semantic loss}\\
\hline
It & 0.0274    &    equal        &0.0330\\
\hline
is & 0.0163   &   rights         & 0.0705\\
\hline
an & 0.0574   &   for         & 0.0167\\
\hline
important & 0.0164   &   all         & 0.0084\\
\hline
step & 0.0544  &   passengers        & 0.0665\\
\hline
towards & 0.0754   &   -         & -\\
\hline
\end{tabular}
\end{table}

\subsection{Overhead Analysis}
Introducing pre-trained language models to recognize the frame importance before transmission inevitably brings incremental cost and delay. 
 In this part, we classify the application scenarios into three categories and discuss the incremental cost and delay separately. To make the classification easily understood, we take the applications of a CCR (Agent $\mathbf{A}$) helping a visually-impaired person (Human $\mathbf{B}$) to complete daily-life tasks as examples.

\subsubsection{Communication content is generated by generative language models like ChatGPT}  A typical example is that Agent $\mathbf{A}$ answers the visually-impaired $\mathbf{B}$'s questions with the help of ChatGPT. In such applications,  $\mathbf{A}$ only needs to modify the prompts input to ChatGPT to simultaneously highlight the important words. Empirical studies in Section V  show that this operation only causes marginal incremental delay, compared to  that of generating the communication content without emphasizing the important words.

\subsubsection{Communication content has already been pre-generated by other sources}  A typical example is that Agent $\mathbf{A}$ reads a book/newspaper to the visually-impaired $\mathbf{B}$. 
In such applications, the communication content can be totally pre-processed offline (i.e., off the transmission from $\mathbf{A}$ to $\mathbf{B}$). In online transmission, the transmitter only needs to conduct priority-based communications according to the frame's importance. 

\subsubsection{Communication content is generated by other sources in real-time}  A typical example is that Agent $\mathbf{A}$ provides real-time voice navigation service for the visually-impaired $\mathbf{B}$. In such applications, the delay and inference cost of processing the content can not be neglected. Let $t_p$ represent the incremental delay of processing a frame. $t_p$ may vary depending on the network latency, the time taken for processing the  request, and the response time of the language models.
Among these factors, the response time of the language models is further determined by the complexity of the input text, the available computational resources, and the load on the server.
The original language models have a huge number of parameters (e.g., BERT has 340 million parameters, and ChatGPT based on GPT 3.5 has 175 billion parameters), which require significant computational resources to perform inference and may introduce a long delay. Despite the substantial cost and delay, it remains worthwhile in some applications for reliable  semantic information delivery. For example, conveying the navigation information to the visually-impaired  $\mathbf{B}$ reliably at the semantic level may avoid serious dangers. 
To minimize the cost and delay,  one can use a self-hosted version of ChatGPT/BERT or other similar language models. In this way, the inference cost and delay only depend on the specific hardware and software configurations.  Besides, researchers have developed smaller and more optimized versions of language models, such as the simplified DistilBERT and MobileBERT models, which have fewer parameters and can be run on less powerful hardware. These models have lower inference cost and can be used in applications with more constrained computational resources, such as mobile devices or embedded systems.

\section{Semantic Importance-Aware Power Allocation}
To minimize the semantic loss, we can design semantic importance-aware priority-based communication strategies in the physical, access, network, transport, or application layers. There are many communication techniques that can be redesigned. In this section, we demonstrate semantic importance-aware power allocation as an example.

Note that resource allocation spans multiple layers in the protocol stack, e.g., the physical and access layers, as shown in Fig. \ref{Transmitter}. In this letter, we present the semantic importance-aware power allocation in the physical layer. It is assumed that all frames go through a Rayleigh fading channel, then the outage probability of the $i$th frame can be expressed as
\begin{equation}
P_i^{out}=P\left(\frac{p_i|h|^2}{\sigma^2}<\gamma_{th}\right)=P\left(|h|^2<\frac{\gamma_{th}\sigma^2}{p_i}\right),
\end{equation}
where $p_i$ is the power allocated to transmit the $i$th frame, $h$ is the channel coefficient obeying the zero mean unit variance complex Gaussian distribution, $\sigma^2$ is the noise variance, and $\gamma_{th}$ is the SNR threshold. 
Let $g=|h|^2$ represent the channel gain. As $h$ obeys a zero mean unit variance complex Gaussian distribution,  $g$ follows an exponential distribution with mean $1$, whose CDF can be expressed as $F_g\left(g\right)=1-e^{-g}$.

Let $g_{i}=\frac{\gamma_{th}\sigma^2}{p_i}$. 
Then, the outage probability of the $i$th frame can be expressed as
\begin{equation}
P_i^{out}=F_g (g_i)=1-e^{-\frac{\gamma_{th}\sigma^2}{p_i}}.
\end{equation}
Hence, the optimization of power allocation to minimize the expected important word errors/semantic loss can be expressed as
\begin{equation}
\label{Problem}
\begin{split}
\min_{\sum_{i=1}^{N}{p_i}=P_{total}}\sum_{i=1}^{N} w_iP_i^{out},
\end{split}
\end{equation}
where $P_{total}$ is the total power of all frames. In ChatGPT-SIAC, $w_i$ denotes the number of important words in the $i$th frame. In BERT-SIAC, $w_i$ is the semantic importance of the $i$th frame defined by the BERT model. Problem (\ref{Problem}) is a typical manifold optimization and we adopt the Manopt toolbox \cite{boumal2014manopt} to solve it. 
For reproducibility, we have released the implementation code in GitHub \cite{GitHub}.

\section{Experiment Results and Discussions}

In this section, we demonstrate our experiment results. 
In the experiments, we evaluate the performance of SIAC on the proceedings of the European Parliament\cite{koehn2005europarl}.
The dataset consists of about $2$ million sentences and $53$ million words.
We pre-process the dataset and generate $100$ batches with each one having $100$ words.
In the experiments, we process a batch at a time. Every $5$ words are packed in a frame and there are a total of $20$ frames to be transmitted at a time. We quantify the importance of each frame by ChatGPT and BERT. All data and experiment details are available on GitHub \cite{GitHub}.

\begin{figure}[htbp]
\centering
\vspace{-0.35cm}
\subfigtopskip=2pt
\subfigbottomskip=2pt
\subfigcapskip=-2pt
\subfigure[Expected important words errors versus the total power, where the word importance is indicated by ChatGPT.]{
\centering \includegraphics[width=0.45\linewidth, keepaspectratio=false]{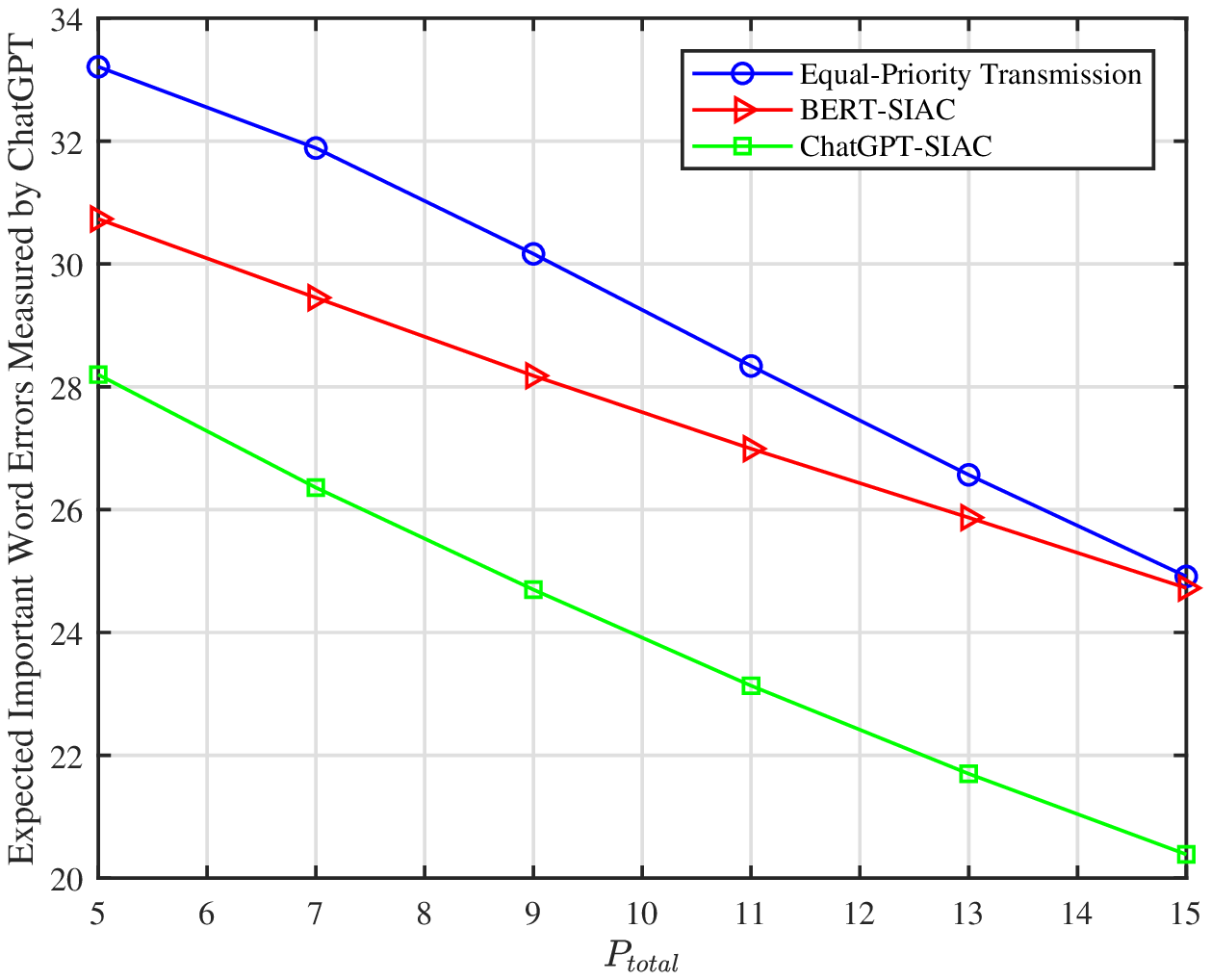}
}
\vspace{0.25cm}
\subfigure[Expected semantic loss versus the total power, where the semantic loss is measured by BERT using (\ref{Loss})]{
\centering \includegraphics[width=0.45\linewidth, keepaspectratio=false]{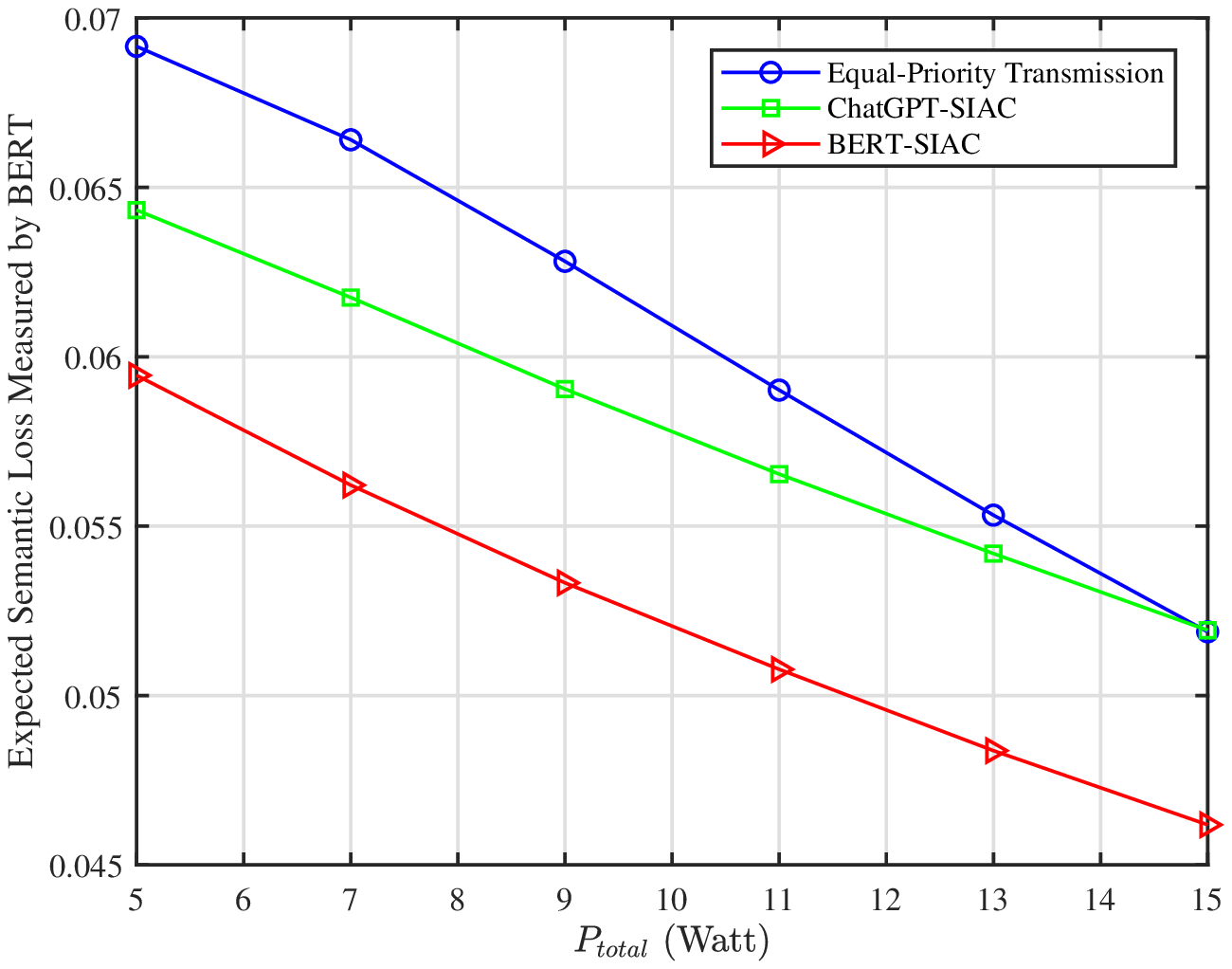}
}
\caption{Comparisons among ChatGPT-SIAC, BERT-SIAC, and equal-priority transmission.}
\label{two graphs}
\end{figure}

In the experiments,  
three schemes are compared, including ChatGPT-SIAC, BERT-SIAC, and equal-priority transmission. 
We set  $\sigma^2=1$ Watt, and $\gamma_{th}=0$ dB. $P_{total}$ is set ranging from $5$ Watt to $15$ Watt.
The power allocation among $20$ frames in a batch is conducted by solving the problem in (\ref{Problem}). We compare three schemes in terms of two performance metrics caused by frame outages. One metric is the expected important word errors. In this evolution, whether a word is important or not is determined by the output of ChatGPT.
The experiment results are  illustrated in Fig. 3(a). As expected, ChatGPT-SIAC achieves optimal performance as its objective is to minimize the expected important word errors. BERT-SIAC also outperforms equal-priority transmission, especially when the transmission power budget is small. The performance gain shrinks as the total power increases.
Another performance metric is the expected semantic loss, which is in (\ref{Loss}) by the BERT model based on semantic similarity.  It is shown in Fig. 3(b) that BERT-SIAC is optimal because its power allocation  is directly proportional to minimizing the expected semantic loss. ChatGPT-SIAC also achieves lower semantic loss than  equal-priority transmission. Comparisons under both performance metrics validate the superiority of the proposed SIAC schemes.

Besides, we conduct an empirical study to investigate the average word generation rate of ChatGPT with different prompts. In the study, we send an original prompt  as well as a copied prompt appending by ``Please highlight the important words to the semantic meanings in every sentence using boldface fonts." to ChatGPT separately. We record the word generation speed in the chat dialogue and illustrate the results in 
Fig. 4(a). Experiment results show that the word generation speed slightly decreases with simultaneously highlighting the important words, indicating marginal delay increment.

Moreover, we also record  the frame importance quantification time in our experiments. The results are illustrated in Fig. 4(b), showing that ChatGPT outputs the frame importance at a fast speed  as it is running on powerful servers. The time consumed by a self-hosted BERT model running on Intel(R) i7-11700@2.5 GHz and NVIDIA GeForce GTX 3090 is relatively higher, which is at a speed of around $25$ frames/minute. This speed is only acceptable for delay-tolerant applications and one can replace the BERT model with the simplified version to reduce the inference delay.

\begin{figure}[htbp]
\centering
\vspace{-0.35cm}
\subfigtopskip=2pt
\subfigbottomskip=2pt
\subfigcapskip=-2pt
\subfigure[Word generation speed of ChatGPT with and without highlighting the important words.]{
\centering \includegraphics[width=0.45\linewidth, keepaspectratio=false]{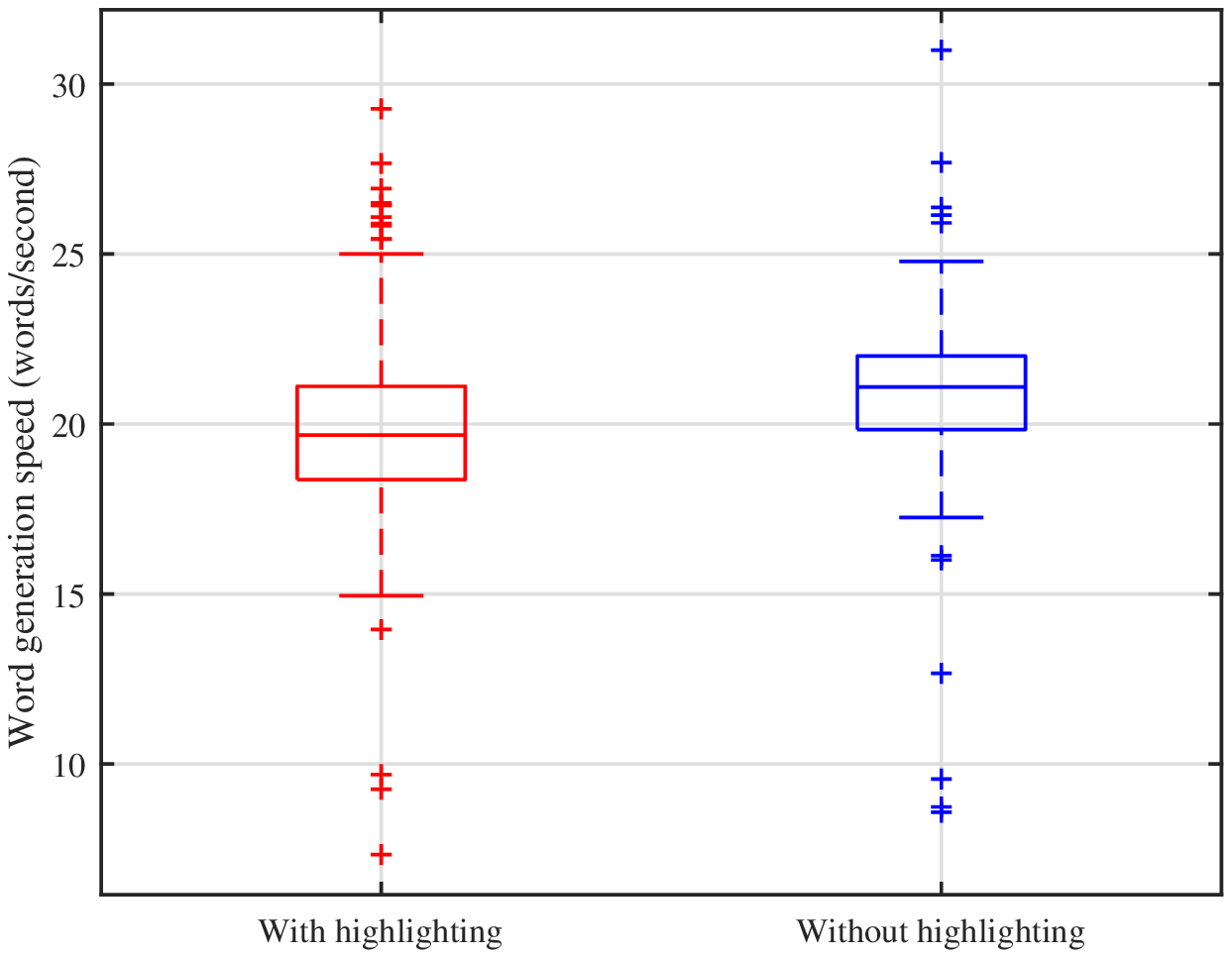}
}
\vspace{0.25cm}
\subfigure[Frame importance quantification time of self-hosted BERT and ChatGPT]{
\centering \includegraphics[width=0.45\linewidth, keepaspectratio=false]{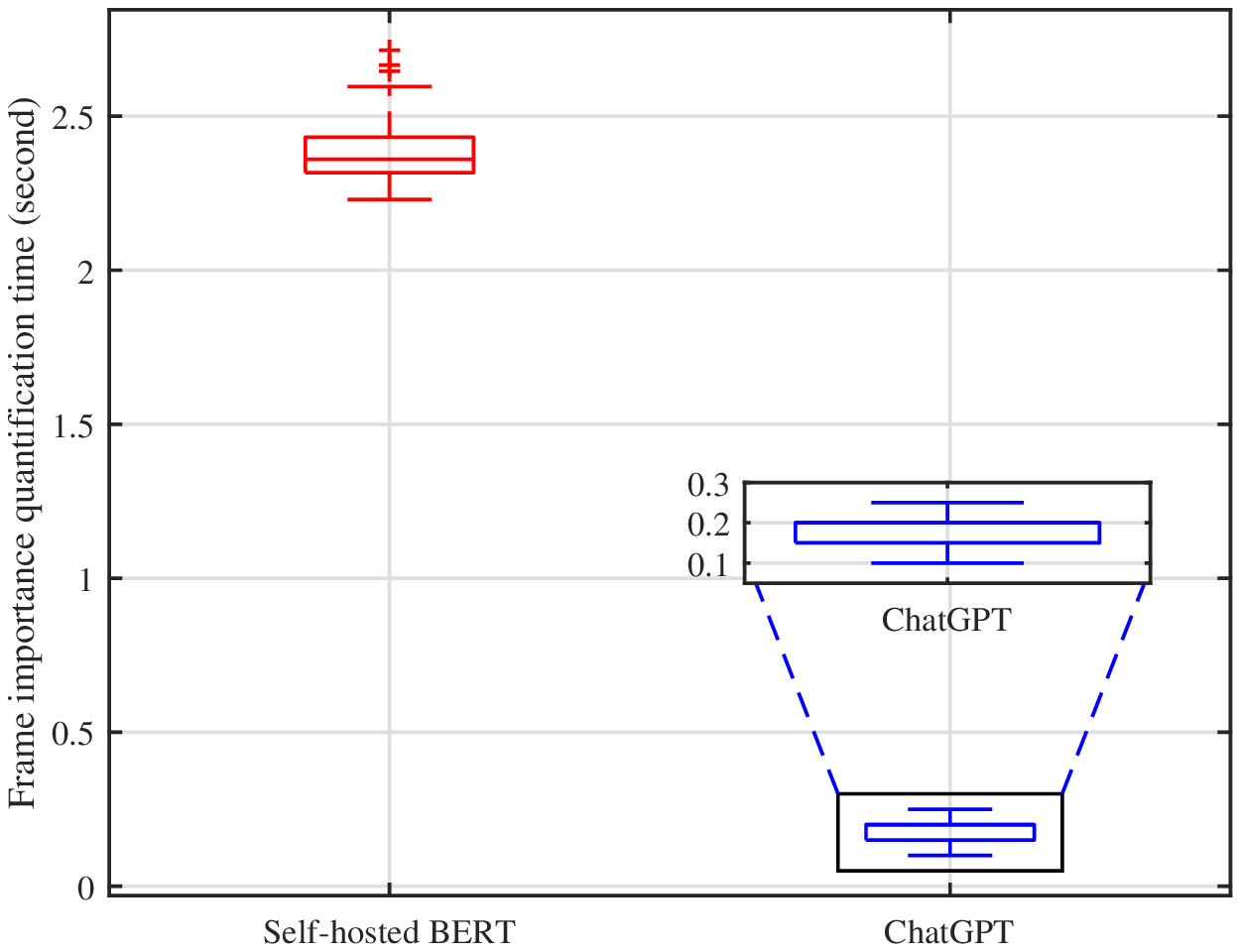}
}
\caption{Empirical study on the word generation speed of ChatGPT, frame importance quantification time of self-hosted BERT and ChatGPT.}
\label{fig:two graphs}
\end{figure}
\section{Conclusions}
This letter introduced pre-trained language models to recognize the importance of frames and proposed the SIAC schemes. Semantic importance-aware power allocation was investigated to demonstrate the advantages of SIAC. It was shown that the proposed schemes can well protect the reliability of semantic communications.

\ifCLASSOPTIONcaptionsoff
  \newpage
\fi

\bibliography{NewCL}
\vspace{-25 pt}

\end{document}